# Polarization-wave propagation as a biophysical mechanism of visual cognition

**Hyun Myung Jang,[1,2]\* Youngwoo Jang,[3] and Hyeon Han[4]**

[1]Research Institute of Advanced Materials, Seoul National University, Seoul 08826, Republic of Korea. [2](Permanent Address) Pohang University of Science and Technology (POSTECH), Pohang 37673, Republic of Korea. [3]Department of Cardiology, Gil Medical Center, Gachon University College of Medicine, Incheon 21565, Republic of Korea. [4]Department of Materials Science and Engineering, Pohang University of Science and Technology (POSTECH), Pohang 37673, Republic of Korea.

\* Correspondence and requests for materials should be addressed to H. M. J. (email: hmjang@postech.ac.kr).

## Abstract

Recent experimental studies indicate that visual cognition is accompanied by slowly propagating biophysical travelling waves in cortical tissue. Here we propose polarization waves as a coherent physical framework for visual cognition. We first compute the propagation of scalar potential fields generated by impressed ionic currents in the primary visual cortex using a telegraph-type model and extract the velocity of the moving potential ridge. By exploiting the linear convolution structure, we then demonstrate that the scalar potential field $\phi(x,t)$ and the polarization wave $P(x,t)$, arising from slowly oscillating neuronal dipoles, propagate with identical velocities. Remarkably, this velocity coincides with the independently predicted propagation speed of the cognitively inferred modulated wave (~1.5 cm/s). Because ionic influx entering a single optic-nerve channel integrates signals from more than a hundred photoreceptors, the resulting polarization field necessarily spans a distribution of wave numbers. We show that amplitudes of such multi-$k$ polarization waves undergo dispersive spreading in time, which possibly suppresses cross-channel interference in visual perception.

# Introduction

Approximately 90% of the information humans acquire from the external world is visual in origin. Yet it remains unclear whether a biological organism composed of roughly 60 trillion cells can directly perceive the physical nature of visible light. The oscillation frequency of incoming optical waves is extremely high: for green light, a typical angular frequency is $\omega \approx 3.5 \times 10^{15}$ s$^{-1}$, corresponding to a photon frequency $f = \omega/2\pi \approx 5.6 \times 10^{14}$ Hz. Green LED spectra exhibit a half-width at half-maximum (HWHM) that is approximately 3% of their central optical frequency[1–3]. Treating the HWHM as an effective uncertainty $\Delta\omega$ in the photon frequency, we obtain $\Delta\omega \approx 0.03\omega \approx 1.1 \times 10^{14}$ s$^{-1}$ ($\Delta f \approx 1.7 \times 10^{13}$ Hz). In contrast, the temporal persistence of a green visual image in humans is only ~40 ms (typically 30–50 ms)[4]. This does determine a characteristic cell-biological frequency $f_{\text{cb}} \approx \frac{1}{0.04 \text{ s}} \approx 25$ Hz, $\omega_{\text{cb}} = 2\pi f_{\text{cb}} \approx 150$ s$^{-1}$. Therefore, the biologically accessible frequency $\omega_{\text{cb}}$ is exceedingly small compared with both the optical frequency and even the uncertainty ($\Delta\omega$) in the optical frequency: $\omega_{\text{cb}} \approx 150 \ll \Delta\omega \approx 1.1 \times 10^{14} < \omega \approx 3.5 \times 10^{15}\ s^{-1}$. Therefore, visible light oscillates far too rapidly for any biological structure to track directly; it cannot serve as a "physically resolved entity" for the nervous system. Instead, visible light interacts with the rapidly fluctuating dipole moments of electronic clouds—whose natural oscillations lie in the same $\sim 10^{15}\ s^{-1}$ range—thereby inducing electronic transitions[5–7].

Because all biological organisms perceive the world through the constraints of their own sensory and neural architectures, a principle often associated with brain-centered cosmology[8-10], humans may interpret internally generated modulations of optical signals as if they reflected properties of the external light itself. In other words, the biological perceptual system inevitably interprets the biologically accessible frequency $\omega_{\text{cb}}$ as the "perceived" frequency of the incoming light. Thus, the visual system effectively interprets external light through a dramatically down-converted modulated wave ($\omega_{\text{cb}} \approx 150\ s^{-1}$), rather than through the high-frequency optical field itself ($\omega \approx 3.5 \times 10^{15}$ s$^{-1}$). Although this low-frequency modulated wave may explain the observed low-frequency travelling waves in the primary visual cortex (V1) cells[11-14], the precise physical nature of the cognitively inferred low-frequency modulated waves remains unexplored.

Here, we present that the cognitively inferred modulated waves result in the slowly propagating polarization wave packet in the visual cortex. (i) We first predict that the propagation velocity of this modulated wave is ~1.5 cm/s, which is similar to or slightly lower than the experimentally observed values ($1.6{\sim}2.2\ cm/s$) of travelling waves in the visual cortex[15,16]. (ii) We then extract the propagation velocity of the scalar potential field generated by impressed ionic currents in V1 by simulating the space-time map using a telegraph-type differential equation. (iii) Exploiting the linear convolution structure[17-19], we show that this scalar field $\phi(x,t)$ and the polarization field $P(x,t)$, arising from slowly oscillating neuronal dipoles, propagate with identical velocities. Remarkably, this velocity agrees well with the independently predicted velocity of the cognitively inferred modulated wave (~1.5 cm/s). (iv) Because the ionic influx entering a single optic-nerve channel integrates signals from more than a hundred photoreceptors[20,21], the resulting polarization field inevitably contains a distribution of wave numbers. We analyze such a multi-$k$ polarization wave and show that its amplitude exhibits dispersive spreading in time. This spreading can play a functional role: by temporally separating signals arriving from adjacent optic pathways, it helps suppress cross-channel interference and thereby stabilizes visual perception. For this mechanism to be effective, the decaying lifetime $\tau_{lf}$ must be comparable to or shorter than the inter-channel time delay.

## Results

**(i) Propagation velocity of the cognitively inferred modulated wave.**

The wave equation for the cognitively inferred modulated wave can be obtained by the superposition of two incident photon waves. This can be illustrated by considering two simultaneously incident photon waves of identical amplitude A but slightly different frequencies, $\omega_o - \Delta\omega$ and $\omega_o + \Delta\omega$ (Fig. 1a). Their superposition yields a modulated wave with an envelope oscillating at $\Delta\omega$, the maximal perceptual frequency difference encoded by the system. Let the displacements of the two sinusoidal waves be $y_1(x,t)$ and $y_2(x,t)$. Their sum becomes $y(x,t) = y_1(x,t) + y_2(x,t) = A(cos\alpha + cos\beta)$, where $\alpha$ and $\beta$ are defined as $\alpha = (\omega_0 + \Delta\omega)t - (k_0 + \Delta k)x$, $\beta = (\omega_0 - \Delta\omega)t - (k_0 - \Delta k)x$. Using the identity $\cos\alpha + \cos\beta = 2\cos\left[\frac{\alpha+\beta}{2}\right]\cdot\cos\left[\frac{\alpha-\beta}{2}\right]$, we obtain

$$y(x,t) = 2A\cos(\omega_0 t - k_0 x)\cos(\Delta\omega t - \Delta k x), \tag{1}$$

as illustrated in Fig. 1b. Here the first cosine represents the carrier wave, whereas the second cosine defines the slowly varying modulation envelope

$$y_{\text{mod}}(x,t) = 2A\cos(\Delta\omega t - \Delta k x). \tag{2}$$

Because the biological perceptual system is constrained by its own temporal resolution, it inevitably interprets the envelope frequency $\Delta\omega$ as the "perceived" frequency of the incoming light. Consequently, we propose

$$\Delta\omega = \omega_{cb} \approx 150 \text{ s}^{-1}. \tag{3}$$

The corresponding spatial modulation scale $\Delta k$ can be estimated from the minimum detectable spatial separation, $\Delta x_{min} \approx 10^{-4}$ m, using the uncertainty-based relation ($\Delta p \Delta x = \hbar \Delta k \Delta x \sim \hbar$):

$$\Delta k = \frac{1}{\Delta x} \approx \frac{1}{\Delta x_{min}} = 10^{+4} \, m^{-1}. \tag{4}$$

However, the group velocity of such a modulated wave reveals that it cannot correspond to the physical propagation of photons along the optic nerve. Using Eq. (2), $v_m = \frac{\Delta\omega}{\Delta k} \approx \frac{150}{10^4} = 1.5 \times 10^{-2} \, (m/s) = 1.5 \, (cm/s)$. This velocity is $\sim 10^{10}$ times slower than the speed of light; indeed, $\frac{v_m}{c} = \frac{1.5}{3 \times 10^{10}} \approx 5 \times 10^{-11}$. If one attempted to interpret this as a slowed electromagnetic wave, the corresponding refractive index of the optic-nerve medium would need to be in the order of $n \sim 2 \times 10^{10}$—an impossibility, since biological tissue in this region has a refractive index of only $\approx 1.4$ (Ref. 22). Therefore, the modulated wave relevant to visual cognition ($\Delta\omega \approx 150$ s$^{-1}$, $\Delta k \approx 10^4$ m$^{-1}$, $v_m \approx 1.5$ cm/s) cannot be a transverse electromagnetic wave; it must be a fundamentally different, bioelectrically mediated phenomenon.

On the basis of the above examinations, we propose the following view: The modulated input wave that underlies visual cognition is not the external optical wave itself but an internally generated representation. This representation may arise either (i) as a perceptual construct created by the cognitive system, or (ii) as a bioelectric scalar wave emerging from the impressed currents injected into the primary visual cortex (V1) cells. Even in the latter case, its propagation velocity (~1.5 cm/s) and quasi-static nature indicate that it is fundamentally distinct from the transverse electromagnetic waves of light. Thus, the modulated wave [Eq. (2)] remains a cognitively inferred representation rather than a real physical entity of the optical stimulus. We will show later that $y_{mod}(x, t)$ effectively plays a role of the bioelectric input wave in the V1 cell layer for visual cognition.

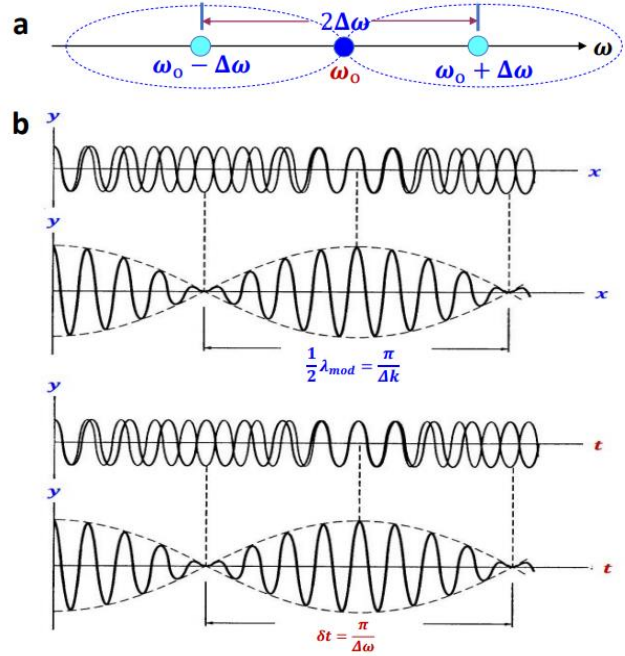

**Fig. 1. Formation of a modulated wave by superposition of two waves.** (**a**) Schematic representation of two simultaneously incident photon waves of identical amplitude but slightly different frequencies, $\omega_o - \Delta\omega$ and $\omega_o + \Delta\omega$. Under this condition, $\Delta\omega$ can be viewed as the maximal perceptual frequency difference and, thus, as the uncertainty limit in the angular frequency $\omega$. (**b**) (Upper Panel) Position ($x$)-dependent representation of a modulated wave or beat pattern formed by the superposition of two waves of nearly equal wavelengths. As indicated in the figure, half-wavelength of the modulated envelope ($\frac{1}{2}\lambda_{mod}$) is equal to $\frac{\pi}{\Delta k}$, where $\Delta k$ denotes the wave number of the modulated wave, $k_{mod}$. (Lower Panel) Time ($t$)-dependent representation of a modulated wave or beat pattern formed by the superposition of two waves of nearly equal angular frequencies, $\omega_o - \Delta\omega$ and $\omega_o + \Delta\omega$. As indicated in the figure, the time ($\delta t$) required for the oscillation of $\pi$ radian in the beat pattern is equal to $\frac{\pi}{\Delta\omega}$. This is because $\delta t/\Delta t \approx \delta t\, \Delta\omega = \pi$ according to Heisenberg's uncertainty relation.

### (ii) Cortical travelling waves and scalar potential fields.

Visual information carried by retinal ganglion cells (RGCs) reaches the V1 cells *via* multiple optical nerve channels arranged in parallel, and the final signals arriving in the V1 cell layer take the form of ionic fluxes (Fig. 2). Representative optical nerve channels include the magnocellular (M) and parvocellular (P) pathways. Because each channel has its own axonal conduction velocity and synaptic delay, the arrival times of visual signals differ measurably across channels. These converging signals in V1 have been reported to constitute the primary source for travelling waves propagating along the cortical surface[11-14]. Sato *et al.*[11], through a systematic review of previous works, argued that the existence of travelling waves that can influence spike responses in V1 is well established. More recently, Benigno *et al.*[12] directly observed travelling waves traversing the entire visual cortex, while Davis *et al.*[13] showed that horizontal cortical connections strongly shape intrinsic travelling waves that regulate sensory-evoked responses and perceptual sensitivity. Although these important studies firmly demonstrate the presence and functional relevance of travelling waves in V1, the precise physical nature of these waves remains largely unexplored. In the present work, we propose— *via* a step-by-step theoretical analysis—that these travelling waves are, in fact, slowly propagating polarization waves $P(x,t)$ moving along the cortical surface. We further show that the cortical scalar potential field $\phi(x,t)$, which is linearly related to $P(x,t)$ through linear convolution, propagates across V1 with the same velocity as $P(x,t)$.

We now consider the physicochemical behavior of the ionic flux (Na$^+$, Ca$^{2+}$) injected into the V1 region. This ion flux generates a large impressed current $\boldsymbol{J}_{\text{imp}}$, and the high ion concentration near the input region at $x=0$ gives rise to a chemical-potential gradient $\nabla \mu$ along the $x$-axis near $x=0$. Writing the local chemical potential as $\mu(x) = \mu_0 + RT \ln C(x)$, we obtain

$$\nabla \mu(x) = \frac{RT}{C(x)} \nabla C(x) = -\frac{RT}{C(x)} \frac{\boldsymbol{J}}{D}, \tag{5}$$

where $D$ is the diffusion coefficient of the ions, $R$ is the gas constant (8.314 J·K$^{-1}$·mol$^{-1}$), and $T$ is the absolute temperature. The last equality in Eq. (5) follows from Fick's first law, $\boldsymbol{J} = -D\nabla C$. Because $\nabla \mu(x)$ carries a negative sign, Eq. (5) implies that the ionic current tends to flow in the $+x$ direction (Fig. 2). The corresponding scalar-potential gradient $\nabla \phi(x)$ induced by this current can be written as $\boldsymbol{J} = \sigma \boldsymbol{E} = -\sigma \nabla \phi(x)$, where $\sigma$ is the effective conductivity. Substituting this expression into Eq. (5) yields $\nabla \mu(x) = \frac{\sigma}{D} \frac{RT}{C(x)} \nabla \phi(x)$. This relation shows that the chemical-potential gradient and the scalar-potential gradient are directly interrelated with each other. Consequently, when ions diffuse from regions of high chemical potential to regions of low chemical potential along the $+x$ direction, the front of the high-$\phi(x)$ region also advances along the $+x$-axis. In other words, the $\phi(x,t)$ field forms a front that moves roughly parallel to the cortical surface.

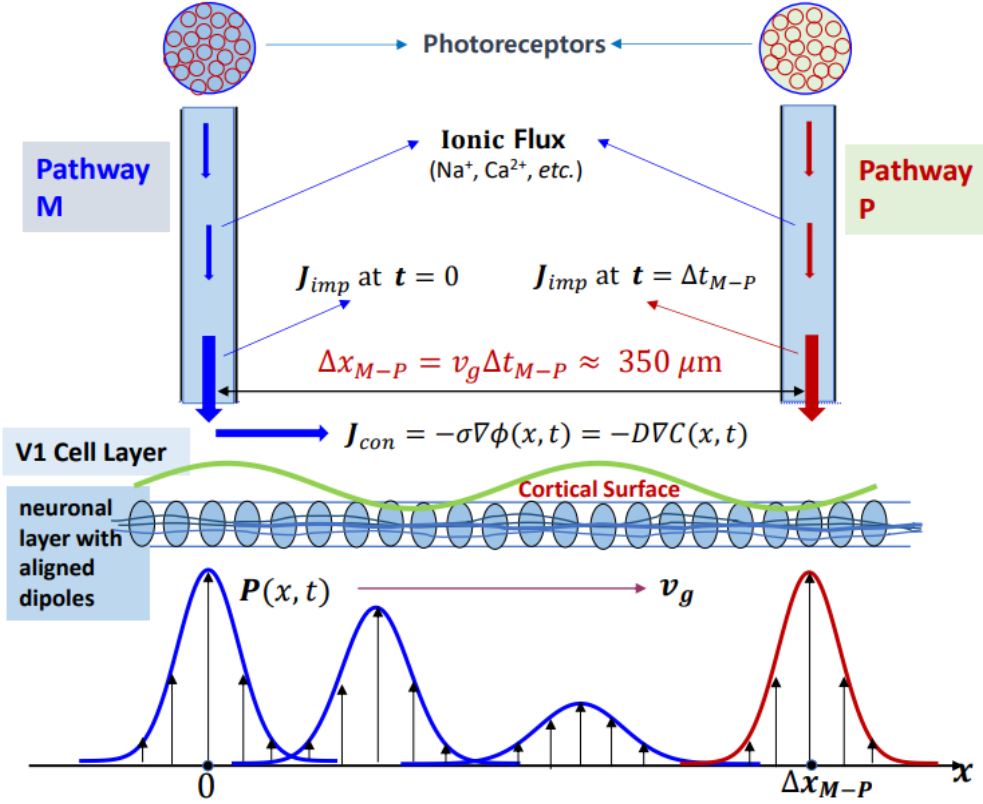

**Fig. 2. Schematic representation of the primary visual cortex (V1) region.** Visual information carried by RGCs reaches V1 region *via* multiple optical nerve channels arranged in parallel. Only two representative nerve channels, M and P, are displayed in this schematic figure. Since $\mathbf{J}_{con} = -\sigma\nabla\phi(x,t)$, the conduction current $\mathbf{J}_{con}$ along $+x$-direction is proportional to $\nabla P(x,t)$ according to the linear convolution theorem. The scalar potential $\phi(x,t)$ is measured on the one-dimensional strip along $+x$-axis located at a constant vertical distance $r_o$ above the neuronal layer in which travelling dipoles are aligned perpendicular to $x$-axis. As marked with blue curves in the bottom panel, the polarization wave packet $P(x,t)$ formed at $x = 0$, *i.e.*, at the exit of M channel (at time $t = 0$) moves along $+x$-axis but gradually spreads with the passage of time due to non-zero $\beta$ [Eq. (13)]. At $t = \Delta t_{M-P}$, a new polarization wave packet is formed at the exit of P channel which is $\Delta x_{M-P}$ away from the exit of M channel. This new wave packet centered at $\Delta x_{M-P}$ is marked with red color. Therefore, we establish the following relation between $\Delta x_{M-P}$ and the inter-channel time delay: $\Delta x_{M-P} = v_g \Delta t_{M-P}$.

Having obtained a qualitative picture of how $\phi(x,t)$ moves, we next consider the inter-channel time delay $\Delta t$ (Ref. 23-25) which is crucially important in stabilizing visual perception. Schmolesky *et al.*[23] report that the time delay between signals in an M channel and those in an adjacent P channel is approximately 23 ms (= 63-40 ms) while Maunsell *et al.*[24] report a broader range of 10–30 ms. We will show that the time delay $\Delta t$ should be larger than the effective lifetime ($t_{lf}$) of the polarization wave packet in V1. If we adopt the reported value by Schmolesky *et al.*[23], we can write the following requirement: $\Delta t = 23 \, ms > t_{lf}$. If this inequality were violated, the residual visual information from the M pathway would mix with newly arriving information from the P pathway, leading to cross-channel interference in visual perception.

**(iii) Dynamics of the scalar potential field and its propagation velocity.**

In this section, we analyze the propagation velocity of the scalar potential field $\phi(x,t)$ in the cortical medium by performing simulations based on the telegraph-type differential equation. We then compare this velocity with the characteristic propagation speed of the cognitively inferred modulated wave, $v_m = \frac{\Delta \omega}{\Delta k} = 1.5$ cm/s. If the two velocities match, this would imply that the scalar potential field in V1 shares the same fundamental frequency $\omega$ and wave number $k$ as the modulated wave $y_{\text{mod}}(x,t)$ relevant to visual cognition. Because $\phi(x,t)$ is linearly related to the cortical polarization field *via* a convolution structure, the potential field $\phi(x,t)$ and the polarization wave are expected to propagate with the same velocity ($v_p$). Therefore, if two velocities ($v_m$ and $v_p$) match, this suggests that there exists an input-response relation between $y_{\text{mod}}(x,t)$ and the polarization wave $P(x,t)$ *via* a linear convolution structure with $y_{\text{mod}}(x,t)$ being the input and $P(x,t)$ as the response. More specifically writing, $P(x,t) = \int \chi(x - x') y_{\text{mod}}(x',t) \, dx'$, where the kernel $\chi(x - x')$ represents dipolar susceptibility. Thus, if the cortical travelling waves in V1 exhibit a speed of approximately $v \approx 1.5$ cm/s, which is nearly identical to $v_m$, one can conclude that the experimentally observed cortical travelling waves[15,16] are, in fact, responding polarization waves and, thus, the observed cortical travelling waves are responding to the input modulated wave which is directly related to visual cognition in V1.

To derive the governing equation for the scalar potential ridge, we combine the condition of charge continuity with the relaxation dynamics of dipole alignment. In an isotropic, homogeneous cortical medium operating in the low-frequency regime, where the high-frequency component of the displacement current can be neglected, the divergence of the total current vanishes, and the total current consists of four contributions: (i) conduction current, (ii) displacement current, (iii) polarization current, and (iv) impressed ionic current generated by neural activity. Thus,

$$\nabla \cdot \boldsymbol{J}_{tot} = 0, \quad \boldsymbol{J}_{tot} = \sigma \boldsymbol{E} + \varepsilon \frac{\partial \boldsymbol{E}}{\partial t} + \frac{\partial \boldsymbol{P}}{\partial t} + \boldsymbol{J}_{imp} \tag{6}$$

Here, $\sigma$ is the tissue conductivity, $\varepsilon$ the dielectric permittivity, $\boldsymbol{P}$ the polarization density, and $\boldsymbol{J}_{imp}$ the impressed ionic current. The electric field is given by $\boldsymbol{E} = -\nabla \phi$. Since the dipolar polarization in cortical tissue responds at low frequency, we adopt a Debye-type relaxation description:

$$\tau_p \frac{\partial \boldsymbol{P}}{\partial t} + \boldsymbol{P} = \varepsilon_0 \chi \boldsymbol{E}, \tag{7}$$

where $\chi$ is the effective dielectric susceptibility, $\varepsilon_0 = 8.854 \times 10^{-12}\, J^{-1}C^2 m^{-1}$ the permittivity of free space, and $\tau_p$ the dipole-alignment relaxation time. Assuming a single $(k, \omega)$-mode for $\boldsymbol{P}(x, t)$ and neglecting phase-lag effects in the low-frequency limit ($e^{i\delta} = 1$), we can eliminate $\boldsymbol{P}$ from Eqs. (6) and (7). After carrying out several algebraic steps (details in Supplementary Information), we obtain a scalar-potential equation which is known as the telegraph-type partial differential equation (PDE) or Goldstein–Kac telegraph equation[26, 27]:

$$\tau \frac{\partial^2 \phi}{\partial t^2} + \frac{\partial \phi}{\partial t} = D \nabla^2 \phi + S(x, t), \tag{8}$$

where the effective relaxation time is defined as $\tau = \tau_p \left(\frac{P_0}{\phi_0}\right)/(2\varepsilon_0 \chi k)$, the effective diffusion coefficient as $D = \frac{\sigma}{C_{\text{eff}}} = \frac{\sigma}{2\varepsilon_0 \chi k^2}$, with $C_{\text{eff}}$ denoting an effective capacitance factor and $k$ the wave number. The source term is given by $S(x, t) = \frac{1}{C_{\text{eff}}} |\nabla \cdot \boldsymbol{J}_{imp}(x, t)|$, which is proportional to the divergence of $\boldsymbol{J}_{imp}$.

Under the single-mode *ansatz*, $\phi(x, t) = \phi_0 e^{-i(kx - \omega t)}$. Substituting this form into Eq. (8) yields the dispersion condition: $-\tau \omega^2 + i\omega + Dk^2 = 0$. In the regime $4\tau D k^2 \gg 1$, the solution becomes

$$\omega = \pm \sqrt{\frac{D}{\tau}}\, k + \frac{i}{2\tau}. \tag{9}$$

Thus, the propagation speed of the scalar potential field is $v = \frac{\omega}{k} = \sqrt{D/\tau}$. Taking the temporal persistence inferred from the visual lifetime as the effective relaxation time ($\tau = 42$ ms) and using the modulated-wave velocity $v_m = 1.5$ cm/s, we obtain an effective diffusion constant $D = v^2\tau \approx 9.45 \times 10^{-6}$ m$^2$/s, consistent with experimentally measured cortical conductivity–permittivity values[28]. For multi-$(k, \omega)$-mode wave packets, the simulated propagation velocity may differ from 1.5 cm/s. Because the damping term $e^{-t/(2\tau)}$ $(= e^{+i\omega t})$ suppresses high-$k$ modes (since $\tau \propto 1/k$), the center of the surviving wave-packet spectrum can shift slightly.

Figure 3a shows the solution of Eq. (8) in 1-D when two optical-channel inputs, separated by 0.4 mm, deliver synaptic ion influxes with an arbitrary temporal delay of 30 ms. The front velocity was estimated by linear regression of the $\phi(x, t)$ ridge in the space–time map. The simulation is grid-convergent: halving $\Delta x$, $\Delta t$ changed $v$ by $< 1\%$. The space–time map exhibits diagonal ridges, whose slope yields a propagation velocity $v = \frac{1.5 \text{ cm}}{\text{s}}$. The nearly linear ridge trajectory indicates that the propagation velocity remains stable over time, confirming a constant-velocity traveling ridge. The central conclusion from this figure is that the scalar-potential ridge propagates at $v = \frac{1.5 \text{ cm}}{\text{s}}$, identical to the velocity of the cognitively relevant modulated wave $v_m = \frac{\Delta \omega}{\Delta k}$. Thus, the cortical scalar-potential field shares the same fundamental $(\omega, k)$-structure as the cognitive modulated wave. By the linear convolution theorem, the polarization-wave ridge must propagate at the same speed. Consequently, the experimentally observed cortical travelling waves with velocities 1.6~2.0 cm/s (Ref. 15) most naturally correspond to polarization waves, and they are responding to the input modulated wave, in accordance with the input-response relation *via* a linear convolution. Figure 3b displays snapshot profiles of $\phi(x, t)$ showing the temporal drift of the peak. From the ridge displacement between $t = 0$ and $t = 50$ ms, the peak-drift velocity is estimated as: $v \approx \frac{0.8 \text{ mm}}{50 \text{ ms}} = 1.6$ cm/s. This estimate aligns closely with: (i) the trajectory-derived value from the space–time map, and (ii) the uncertainty-based prediction $v_m = 1.5$ cm/s. Thus, both methods confirm a robust propagation velocity of $v \approx 1.5 – 1.6$ cm/s.

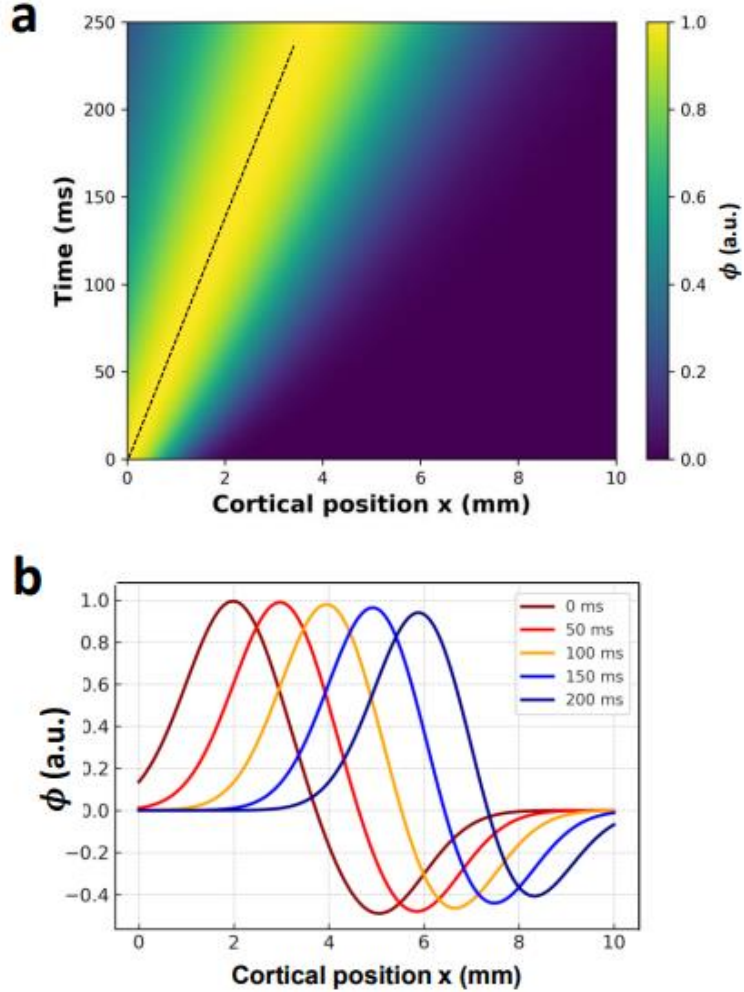

**Fig. 3. Cortical scalar potentials simulated using the telegraph-type equation. (a)** Space-time map of the cortical scalar potential field $\phi(x,t)$ simulated using the telegraph-type partial differential equation, Eq. (8). The nearly linear ridge trajectory is marked with a black broken line. The propagation velocity of scalar potential ridge estimated using this trajectory line is $v = 1.5 \ cm/s$. **(b)** Snapshot profiles of $\phi(x,t)$ at various times, showing the temporal drift of the peak. The propagation velocity estimated using the peak values of $\phi(x,t)$ is approximately equal to the propagation velocity of $\phi(x,t)$ field obtained from the trajectory line in the above space-time map.

**(iv) Linear convolution relation between scalar potential and polarization.**

Let us consider a one-dimensional strip along the *x*-axis at a fixed observation depth $z_0$, located at a constant vertical distance $r_0$ above the neuronal layer in which dipoles are aligned (Fig. 2). The scalar potential measured on this strip, $\phi(x,t)$, is related to the polarization wave $P(x,t)$, which travels along the *x*-direction with the propagation velocity $v$, through a linear convolution. This convolution relation can be written as

$$\phi(x,t;z_0) = \int K(x-x';z_0)\, P(x',t)\, dx' = (K*P)(x-vt) \tag{10}$$

Eq. (10) expresses velocity preservation: although the kernel $K$ may smooth the waveform, it does not alter the phase or propagation speed of the travelling polarization wave. Thus, $\phi(x,t)$ inherits the same front velocity as $P(x,t)$. Applying forward Fourier transform to Eq. (10) yields the linear convolution relation in $k$-space:

$$\phi(k,t) = K(k)P(k,t). \tag{11}$$

Here $P(k,t) = \int P(x',t)e^{+ikx'}dx'$, $K(k) = \int K(x-x')e^{+ik(x-x')}dx$.

If the polarization wave consists of a single $k$-mode, its Fourier transform is $P(k,t) = \int P_0 e^{-i(k_0 x' - \omega t)} e^{+ikx'} dx' = 2\pi P_0 e^{+i\omega t} \delta(k_o - k)$. Substituting this relation into Eq. (11) gives $\phi(k,t)$, and inverse Fourier transformation yields $\phi(x,t) = \int \phi(k,t)e^{-ikx}dk = \phi_0 e^{-i(k_0 x - \omega t)}$, where $\phi_0 = 2\pi P_0 K(k_0)$. Thus, the amplitude of the scalar potential ($\phi_0$) depends solely on the value of the kernel at the carrier wave number, $K(k_0)$. Consequently, the peak propagation speed of $\phi(x,t)$ is identical to that of the polarization wave and is given by $v = \omega/k_o$. For a translationally invariant single-mode wave, application of the convolution theorem yields

$$(K * e^{-ik_0 x})(x) = e^{-ik_0 x} \int K(x-x') e^{+ik_0(x-x')} dx = K(k_0) e^{-ik_0 x} \tag{12}$$

$K(k_0)$ was obtained by using the definition of $K(k)$ given in Eq. (11). The right-hand side is equal to $(2\pi P_0 e^{+i\omega t})^{-1} \phi(x,t)$, demonstrating that any kernel is permissible for $\phi(x,t)$ to be a travelling wave as long as its Fourier transform satisfies $K(k_0) \neq 0$. In our case, we have already shown that $K(k_0) = \phi_0/(2\pi P_0)$.

As noted earlier, each optical nerve channel receives ion-influx signals originating from more than 100 photoreceptors, representing diverse wavelengths and temporal patterns[20,21]. Thus, the V1 travelling wave is expected to reflect multiple $(k,\omega)$ components, not a single mode. In such a case, the polarization wave packet may be written as $P(k,t) = P(k)e^{+i\omega(k)t} =$ $P_0 e^{-\alpha(k-k_0)^2} e^{+i\omega(k)t}$, where $\alpha = 1/(2\sigma^2)$ and $\sigma$ is the standard deviation of the $k$-distribution. The scalar potential satisfies an analogous form: $\phi(k,t) = \phi_0 e^{-\alpha(k-k_0)^2} e^{+i\omega(k)t}$. Substituting these expressions into Eq. (11) shows that $K(k) = \phi_0/P_0 \neq 0$, which is not restricted to $k = k_o$. Therefore, the scalar potential field $\phi(x,t)$ is also a wave packet moving with the same velocity as $P(x,t)$ for all possible values of the wave number.

To determine the propagation speed of the wave-packet peak, consider Taylor expansion of $\omega(k)$ about $k_0$ (Ref. 29):

$$\omega(k) = \omega(k_0) + (k-k_0)\left(\frac{d\omega}{dk}\right)_{k_0} + \frac{1}{2}(k-k_0)^2 \left(\frac{d^2\omega}{dk^2}\right)_{k_0}$$

$$= \omega_0 + (k-k_0)v_g + \frac{1}{2}(k-k_0)^2 \beta, \tag{13}$$

where $v_g = (d\omega/dk)_{k_0}$ is the group velocity, $\beta = \left(\frac{d^2\omega}{dk^2}\right)_{k_0}$ is the dispersion coefficient. Using Eq. (13) in the expression for $P(k,t)$ and performing the inverse Fourier transform yields $P(x,t) = P(k_0)e^{-i(k_0 x - \omega_0 t)} A(t) \exp\left[-\frac{(x-v_g t)^2}{4\left(\alpha - \frac{i\beta t}{2}\right)}\right]$, where $A(t)$ is a complex prefactor. An identical result is obtained for $\phi(x,t)$. The exponential envelope is maximized when $x = v_g t$, showing that both the polarization and scalar-potential wave packets propagate with the same group velocity $v_g$. This also supports our estimate of $v \approx 1.6\ cm/s$ from the peak-drift velocity of $\phi(x,t)$ profile presented in Fig. 3b.

**(v) Thermodynamic driving force of slowly moving polarization waves.**

Up to this point, we have not explicitly identified the thermodynamic mechanism that drives the slow tangential motion of the polarization wave $P(x,t)$. We argued qualitatively that ionic diffusion from regions of high to low chemical potential would produce a scalar-potential front propagating along the +x-axis. We now build an explanation for how this gradient also drives the polarization wave. From Fick's law and Ohm's law, concentration and potential gradients satisfy $D\nabla C(x,t) = \sigma \nabla \phi(x,t)$. We previously showed that $P(x,t)$ is mediated by $\phi(x,t)$ according to the relation that $\nabla P(x,t) = \frac{1}{2\pi K(k_0)} \nabla \phi(x,t)$. Substituting this into the first expression gives

$$\nabla P(x,t) = \frac{1}{2\pi K(k_0)}\left(\frac{D}{\sigma}\right) \nabla C(x,t) = \frac{1}{2\pi K(k_0)}\left(\frac{b}{\sigma}\right) C(x,t) \nabla \mu(x,t), \tag{14}$$

where $\nabla\mu = (RT/C)\nabla C$ and $b$ denotes ionic mobility. The first form of Eq. (14) shows that polarization gradients are directly proportional to concentration gradients. Thus, a spatially varying ionic concentration $\nabla C(x,t)$ produces a polarization gradient, which in turn generates a conduction current $\boldsymbol{J}_{con} \propto -\nabla P$. We can explain more quantitatively by rewriting the first expression of Eq. (14) in the following form: $\nabla P(x,t) = \frac{1}{2\pi K(k_0)}\left(-\frac{1}{\sigma}\right)\boldsymbol{J}_{con}$. We used Fick's first law to obtain this expression. According to this relation, the tangential conduction current, $\boldsymbol{J}_{con} \propto -\sigma\nabla P(x,t)$, is maximal near the propagating front. This region of maximal current moves together with the polarization-wave front at the same propagation speed (Fig. 3b). When the dominant $(k_0, \omega_0)$-mode is considered, the resulting polarization-wave front moves with the velocity $v = \frac{\omega_0}{k_0}$. As ions diffuse from the channel exit at $x = 0$, $\nabla C(x,t)$ gradually decreases, causing the polarization gradient—and hence the driving force of the traveling wave—to diminish over time (Fig. 2).

The second form of Eq. (14) comes from expressing the tangential conduction current using chemical potential[30,31]: $\boldsymbol{J}_{con} = -D\nabla C(x,t) = -bC(x,t)\nabla\mu(x,t)$, with $b$ denoting ionic mobility. By eliminating $\nabla C$ and applying $\frac{P_0}{\phi_0} = \frac{1}{2\pi K(k_0)}$, we obtain Eq. (14). Because $\nabla\mu$ has the dimensions of force, chemical-potential gradients represent the ultimate thermodynamic driving force of polarization-wave propagation. We now find the mobility $b$ in terms of well-known physical parameters. Combining $\nabla\mu(x,t) = \left(\frac{\sigma}{D}\right)\frac{RT}{C(x,t)}\nabla\phi(x,t)$ and $\nabla\phi(x,t) = \left(\frac{\phi_0}{P_0}\right)\nabla P(x,t)$, yields an expression of $\nabla P(x,t)$ in terms of $\nabla\mu(x,t)$: $\nabla P(x,t) = \left(\frac{D}{\sigma}\right)\left(\frac{1}{RT}\right)\frac{1}{2\pi K(k_0)}C(x,t)\nabla\mu(x,t)$. Comparing this with Eq. (14) shows that ionic mobility satisfies $b = \frac{D}{RT}$, which matches the well-known Einstein relation[32]. These arguments extend naturally to polarization wave packets involving many $(k,\omega)$ components (Supplementary Information).

### (vi) Polarization wave packet in the V1 region and its spreading.

As discussed previously, the slowly propagating polarization field $P(x,t)$ in the V1 region is more realistically represented not by a single $(k,\omega)$-mode, but by a wave packet composed of multiple modes distributed around the cognitively fundamental modulation parameters $\Delta\omega$ and $\Delta k$, as defined in Eq. (3) and Eq. (4), respectively. For this polarization wave packet, we obtain the following expression (Supplementary Information for derivation):

$$P(x,t) = P_0 e^{-i(k_0 x - \omega_0 t)} A(t) \exp\left[-\frac{(x-v_g t)^2}{4(\alpha - i\beta t/2)}\right] \quad (15)$$

Here, the amplitude prefactor $A(t)$ is defined as $\{A(t)\}^2 = \left(\frac{\pi}{\alpha - i\beta t/2}\right)$, and $\alpha = 1/(2\sigma^2)$, where $\sigma$ is the standard deviation of the wave-number distribution. It should be noted that $k$-dependent $\omega$ values expressed in Eq. (13) are now smeared in $v_g$ and $\beta$ terms in Eq. (15). Taking the modulus of Eq. (15), the $x$-dependent envelope of the polarization wave packet becomes:

$$|P(x,t)| = P_0 \left|e^{-i(k_0 x - \omega_0 t)}\right| \left(\frac{\pi^2}{\alpha^2 + \beta^2 t^2/4}\right)^{1/4} \exp\left[-\frac{\alpha(x-v_g t)^2}{4(\alpha^2 + \beta^2 t^2/4)}\right] \quad (16)$$

Because $\left|e^{-i(k_0 x - \omega_0 t)}\right| = 1$, it is retained only to indicate that this wave packet originates from the fundamental modulation wave with the basic frequency $\omega_0 = \Delta\omega = \omega_{cb}$ and the basic wave number $k_0 = \Delta k$. Eq. (16) shows that $|P(x,t)|$ has a Gaussian envelope centered at $x = v_g t$, with a width that gradually increases over time, that is, a dispersive spreading characteristic. Importantly, the parameters $\omega_0$ and $k_0$ in Eqs. (15) and (16) do not refer to the physical optical frequency $\sim 3.5 \times 10^{15}\,\text{s}^{-1}$ or optical wave number $\sim 10^7\,\text{m}^{-1}$, but instead correspond to the much smaller modulation parameters $\Delta\omega \approx 150\,\text{s}^{-1}$ and $\Delta k \approx 10^4\,\text{m}^{-1}$. Thus, the wave packet $P(x,t)$ represents the superposition of polarization modes around this cognitively inferred modulation wave, rather than the physical optical field itself.

Because of dispersive spreading, the height (peak amplitude) of the polarization wave packet decreases, and its width increases over time (Fig. 4). To quantify this, let the peak intensity at time $t = 0$ and position $x = 0$ be denoted: $I_{0,0}$. Similarly, let the peak intensity at time $t = \Delta t$ and position $x = \Delta x$ be $I_{\Delta x, \Delta t}$. Using Eq. (16), the ratio of these two intensities becomes:

$$\frac{I_{\Delta x, \Delta t}}{I_{0,0}} = \left(\frac{|P(\Delta x, \Delta t)|}{|P(0,0)|}\right)^2 = \left(\frac{\alpha^2}{\alpha^2 + \beta^2 (\Delta t)^2/4}\right)^{1/2}. \quad (17)$$

Since $\Delta x = v_g \Delta t$, the intensity decays entirely as a function of time. If we define the lifetime of the polarization wave packet as the time at which the peak intensity decays to one-tenth of its initial value, we denote this time as $t_{lf(p)}$. From Eq. (17), one then obtains: $t_{lf(p)} = \frac{(2\sqrt{99})\alpha}{\beta} \approx$ $\frac{20\alpha}{\beta}$. Here, $\sigma = 1/\sqrt{2\alpha}$ is the standard deviation of the wave-number distribution; thus, larger $\sigma$ (broader $k$-distribution) leads to a shorter lifetime $t_{lf(p)}$. The parameter $\beta = (d^2\omega/dk^2)_{k_0}$ is the principal factor controlling the spreading of the wave packet; smaller $\beta$ yields a longer lifetime.

In addition to amplitude decay, the lifetime can also be defined through the spreading of the half-width at half-maximum (HWHM) of the Gaussian envelope. Let the HWHM at $t = 0$ be $W_d(0)$. Similarly, let the HWHM at $t = \Delta t$ be $W_d(\Delta t)$. Then one obtains the following ratio (Supplementary Information for derivation):

$$\frac{W_d(\Delta t)}{W_d(0)} = \left(1 + \frac{\beta^2(\Delta t)^2}{4\alpha^2}\right)^{1/2}. \tag{18}$$

If the lifetime is defined as the time when the HWHM becomes ten times broader than its initial value, one obtains: $t_{lf(w)} = \frac{(2\sqrt{99})\alpha}{\beta} \approx \frac{20\alpha}{\beta}$. The remarkable fact that two distinct definitions of lifetime yield exactly the same functional form $t_{lf} = (2\sqrt{99})\alpha/\beta$ highlights the internal mathematical consistency of the spreading behavior of $|P(x,t)|$ in the V1 region.

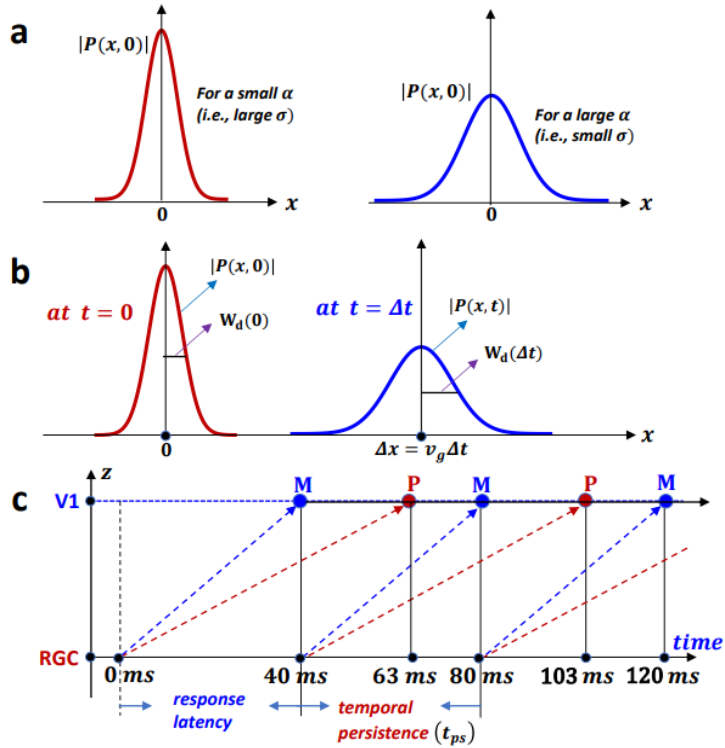

**Fig. 4. Polarization wave packets and temporal persistence.** (a) Effect of the standard deviation on the shape of $|P(x,t)|$ at a fixed time, $t = 0$. Here, $\sigma$ signifies the standard deviation in the wave-number distribution. In other words, $k$-dependent polarization in Gaussian approximation is given by $P(k) = P_o e^{-\alpha(k-k_o)^2}$, where $\alpha = 1/2\sigma^2$. From Eq. (16), it can be shown immediately that $|P(x,0)| = \left(\frac{\pi}{\alpha}\right)^{\frac{1}{2}} P_o \exp\left(-\frac{x^2}{4\alpha}\right)$. It should be noted that the parameter $\alpha$ is now in the denominator side of a Gaussian envelope. In contrast, $\alpha$ is placed in the numerator side of $k$-dependent Gaussian polarization $P(k)$. (b) Propagation of Gaussian wave packet $|P(x,t)|$ with the group velocity $v_g$, showing spreading behavior owing to non-zero $\beta$. Initially, the peak value of $|P(x,0)|$ at $x = 0$ is given by $|P(0,0)| = \left(\frac{\pi}{\alpha}\right)^{\frac{1}{2}} P_o$ with its HWHM denoted by $W_d(0) = \sqrt{4\alpha \cdot \ln 2}$. On the contrary, the peak value of $|P(x,t)|$ at $t = \Delta t$ is given by $|P(\Delta x, \Delta t)| = P_o \left(\frac{\pi^2}{\alpha^2 + \frac{\beta^2(\Delta t)^2}{4}}\right)^{\frac{1}{4}}$, where $\Delta x = v_g \Delta t$. Here, it should be noted that $\Delta x$ is not uncertainty in the position, but denotes the peak position at $t = \Delta t$. It can be shown that $W_d(\Delta t)$ can be expressed in terms of $W_d(0)$ using the following expression: $W_d(\Delta t) = W_d(0)\left(1 + \frac{\beta^2(\Delta t)^2}{4\alpha^2}\right)^{\frac{1}{2}}$ (Supplementary Information for details). (c) A schematic diagram that shows the response latency of 40 ms for the M pathway and the temporal persistence ($t_{ps}$) of the same period of 40 ms. Here, two optical nerve channels, M and P, are arranged in parallel to the $z$-direction.

## Discussion

**(i) Temporal persistence and lifetime of the polarization wave packet.**

In visual cognition, response latency refers to the time required for signals from retinal ganglion cells (RGCs) to reach the primary visual cortex (V1). According to the measurements of Schmolesky et al.[23], the response latency of the M pathway in macaque monkeys is 40 ms, whereas the latency of the neighboring P pathway is about 63 ms. The latter value lies between the arrival times of successive ion-influx signals through the M pathway at about 40 ms and 80 ms (Fig. 4c). This implies that, between 40 ms and 80 ms, visual information obtained at time $t = 0$ from the photoreceptors remains present in the V1 region without being completely extinguished. By the same reasoning, between 80 ms and 120 ms, visual information acquired at $t = 40$ ms is expected to persist in the V1 region (Fig. 4c). These observations lead to the important conclusion that the temporal persistence of a visual image is determined primarily by the response latency of the M pathway, yielding a characteristic value of about 40 ms for macaque monkeys. Although the P pathway exhibits a longer response latency (63 ms), it does not set the temporal persistence; even when referenced to the P pathway, the temporal persistence remains 40 ms, as given by $(103-63)\ ms$ (Fig. 4c).

Let the temporal persistence be denoted by $t_{ps}$. It may then be decomposed into two inter-channel time delays: $t_{ps} = \Delta t_{M-P} + \Delta t_{P-M} = (63 - 40)$ ms $+ (80 - 63)$ ms $= 23$ ms $+ 17$ ms. Here, $\Delta t_{M-P}$ represents the delay of ion-influx arrival through the P pathway relative to the neighboring M pathway (23 ms; Fig. 4c), while $\Delta t_{P-M}$ corresponds to the time interval (17 ms) after the P-pathway activation until the arrival of a new ion-influx signal through the next M pathway (Fig. 4c). The delay $\Delta t_{M-P}$ must be greater than or comparable to the effective lifetime of the polarization wave packet $P(x,t)$, $t_{lf} \approx \frac{20\ \alpha_{(M)}}{\beta}$, as discussed previously. Thus, the condition $\Delta t_{M-P} = 23$ ms $\geq \frac{20\ \alpha_{(M)}}{\beta}$ must be satisfied. If this inequality were violated, the residual visual information from the M pathway would mix with newly arriving information from the P pathway, leading to ambiguity or confusion in visual perception. A similar requirement applies to the delay $\Delta t_{P-M} = 17$ ms, $\Delta t_{P-M} \geq \frac{20\ \alpha_{(P)}}{\beta}$, where $\alpha_{(P)}$ characterizes the polarization wave packet induced by the P-pathway ion influx. Assuming that the dispersion parameter $\beta$ is pathway-independent, one obtains the inequality $\alpha_{(M)} > \alpha_{(P)}$. Finally, it should be noted that the present discussion is based on the experimental data obtained from macaque monkeys[23]. Whether the same quantitative relationships apply directly to human visual cognition remains an open question and requires further experimental validation.

**(ii) Modulated waves *versus* cortical travelling waves.**

As mentioned previously, experimentally observed traveling waves in the visual cortex propagate with velocities in the range of approximately 1.6~2.2 cm/s[15,16]. These values are slightly higher than the characteristic propagation velocity of the cognitively inferred modulated wave, $v_m = \frac{\Delta \omega}{\Delta k} = \frac{\omega_{cb}}{\Delta k} \approx 1.5$ cm/s. Here, it should be noted that the modulated wave described by Eq. (2), with $\Delta \omega = \omega_0$ and $\Delta k = k_0$, can be regarded as a single $(k_o, \omega_o)$-mode wave. The fact that the experimentally observed traveling waves propagate faster than this characteristic single-mode velocity ($v_m = \frac{\omega_0}{k_0} = 1.5$ cm/s) suggests that the cortical traveling waves are not single-mode excitations, but rather polarization wave packets composed of a finite range of wave numbers and frequencies. As described previously, ion-influx signals arriving through a given optic nerve channel originate from information integrated over more than one hundred photoreceptors, each sensitive to a range of wavelengths and frequencies[20,21]. It is therefore natural to expect that the resulting traveling activity in V1 reflects a superposition of multiple $(k, \omega)$ components and exhibits wave-packet–like behavior. For such a wave packet, the relevant propagation speed is the group velocity, $v_g = \left(\frac{d\omega}{dk}\right)_{k_0}$, which generally differs from the phase velocity of the cognitively inferred modulated wave having a single mode character, $v_m = \frac{\omega_0}{k_0} \approx 1.5 \ cm/s$.

The observed enhanced velocity of traveling waves in the visual cortex can also be explained in terms of the damping effect without invoking a multi $(k, \omega)$-mode polarization wave. The damping term $e^{-t/(2\tau)}$, which arises from the imaginary part of the angular frequency $\omega$ [Eq. (9)] through $e^{+i\omega t} = e^{+i(+i/2\tau)t}$, tends to suppress high-$k$ modes since the effective relaxation time $\tau$ is inversely proportional to $k$ [See Eq. (8) for the definition of $\tau$]. This reduced $\tau$ values for high-$k$ modes enhance the damping effect in $P(x,t)$ ($\propto e^{-t/(2\tau)}$) and then leads to an enhanced propagation velocity as $v = \frac{\omega}{k} = \sqrt{\frac{D}{\tau}}$ [Eq. (9)]. In the case of awake monkeys[33], the cortical traveling-wave velocity is an order of magnitude faster than the velocity of the cognitively inferred modulated wave for humans ($v_m \approx 1.5 \ cm/s$). This pronounced difference presumably indicates that the $(\Delta\omega/\Delta k)$-ratio of awake monkeys is much bigger than that of humans.

## Conclusions

The following conclusions are obtained through our systematic study on the slowly propagating biophysical travelling waves in the visual cortex: **(i)** The velocity of the scalar potential ridge, as obtained by using a telegraph-type model simulation, is about 1.5 cm/s. Remarkably, this value is equal to the independently predicted velocity of the cognitively inferred modulated wave ($v_m$ =1.5 cm/s). **(ii)** By exploiting the linear convolution structure, we show that the scalar potential field $\phi(x,t)$ and the polarization wave $P(x,t)$ propagate with identical velocities. **(iii)** We propose that the observed cortical travelling waves are, in fact, the polarization waves that are responding to the cognitively inferred modulated waves, in accordance with the input-response relation in a convolution framework. **(iv)** The polarization field in the V1 region contains a distribution of wave numbers. Such a multi-$k$ polarization wave packet exhibits dispersive spreading in time, which possibly suppresses cross-channel interference and thereby stabilizes visual perception.

## Methods

### (i) Simulations of cortical scalar potentials by telegraph-type equation.

We model the extracellular scalar potential $\phi(x,t)$ in an isotropic, homogeneous cortical tissue under quasi-static conditions by a damped telegraph-type PDE derived from the current continuity and the Debye-type polarization relaxation. The relevant PDE is given in Eq. (8), namely, $\tau \frac{\partial^2 \phi}{\partial t^2} + \frac{\partial \phi}{\partial t} = D \nabla^2 \phi + S(x,t)$. Here $D$ is the effective diffusivity, $\tau$ the effective relaxation time (dominated by polarization/membrane RC dynamics), and $S(x,t)$ the impressed source density (divergence of synaptic ionic currents). Baseline parameters were set to match the independently inferred wave speed $v = \sqrt{D/\tau} \approx 1.5$ cm s$^{-1}$ and persistence $\tau \approx 0.042$ s from psycho-physical constraints. The spatial observation depth along z was held constant (fixed recording plane). A one-dimensional tangential strip of the cortex of length $L = 10$ mm was simulated with a uniform spacing $\Delta x = 50$ μm (grid points $N_x = L/\Delta x + 1$). Time integration covered $T = 250$ ms with $\Delta t = 0.02$ ms (ensuring numerical stability for the chosen explicit scheme; empirically $\Delta t \lesssim 0.1 \sqrt{\Delta x^2/D}$ worked robustly). Two localized thalamo-cortical inputs for 'two-channel exits' were modeled as separable Gaussians in space times ($\otimes$) causal pulses in time: $S(x,t) = A_1 \exp\left[-\frac{(x-x_1)^2}{2\sigma_x^2}\right] p(t-t_1) + A_2 \exp\left[-\frac{(x-x_2)^2}{2\sigma_x^2}\right] p(t-t_2)$, with $x_2 - x_1 = \Delta x_{\text{src}} = 0.4$ mm, $\sigma_x = 0.15$ mm. The temporal kernel $p(\cdot)$ was a brief, causal log-normal–like pulse (rise-and-decay) of the width $w \approx 8$ ms, peaking shortly after its onset; the second source was delayed by $\Delta t_{\text{src}} = 30$ ms to reflect pathway timing differences (M vs. P, synaptic hops). Amplitudes $A_1, A_2$ were equal unless stated.

We used second-order centered differences in space and a damped leap-frog–type update for time: $\phi_j^{n+1} = 2\phi_j^n - \phi_j^{n-1} + \frac{\Delta t^2}{\tau}\left(D\,\delta_{xx}\phi_j^n + S_j^n\right) - \frac{\Delta t}{\tau}\left(\phi_j^n - \phi_j^{n-1}\right)$ with $\delta_{xx}\phi_j^n = \phi_{j+1}^n - 2\phi_j^n + \phi_{j-1}^n$. Neumann boundaries (zero-flux) were imposed by copying edge-adjacent values at each step. Initial conditions are $\phi^0 = \phi^{-1} = 0$. We recorded (i) a space–time map of $\phi(x,t)$ (down-sampled in $t$) and (ii) spatial snapshots $\phi(x,t_k)$ at selected times. The front velocity was estimated by linear regression of the ridge in the space–time map, recovering $v \approx$ 1.5 cm s$^{-1}$ within a few percent. Parameter sweeps varying $\tau$ (0.02–0.06 s) and $D$ accordingly ($v$ fixed) confirmed qualitative robustness of damped traveling ridges, with temporal damping scale $\approx 2\tau$ and spatial attenuation $\ell \approx 2\sqrt{D\tau} \sim 1.3$ mm. All simulations were run in double precision. The method is grid-convergent (halving $\Delta x, \Delta t$ changed $v$ by < 1%).

### (ii) Kernels $K(x)$ and $K(k, \omega)$ for telegraph-type equation.

For a translationally invariant single-mode wave, application of the convolution theorem yields the following equality: $(K * e^{-ik_0 x})(x) = K(k_0)e^{-ik_0 x} = (2\pi P_0 e^{+i\omega t})^{-1}\phi(x,t)$, as previously shown in Eq. (12). This indicates that any kernel $K(x)$ is permissible as long as its Fourier transform satisfies $K(k_0) \neq 0$. In our case, $K(k_0) = \phi_0/(2\pi P_0)$ as $\phi(x,t)$ can be represented by the following relation for a single $(k_o, \omega)$-mode: $\phi(x,t) = \phi_0 e^{-i(k_0 x - \omega t)}$. It is interesting to identify the functional form of $K(k_0)$ for the telegraph-type equation [Eq. (8)] used to simulate the propagation velocity of the scalar-potential front $\phi(x,t)$. Substituting $\phi(x,t) = \phi_0 e^{-i(kx-\omega t)}$ into Eq. (8) yields the following Helmholtz-type equation for $\phi(x,\omega)$ in the $\omega$-domain:

$$(\nabla^2 + k_{eff}^2)\phi(x,\omega) = -\frac{S(x,\omega)}{D} \qquad (19)$$

where $k_{eff}^2 = \frac{\tau\omega^2 - i\omega}{D}$ and $\phi(x,\omega) = \int \phi(x,t)e^{-i\omega t}dt$. Suppose we could find a function $G(x)$ that solves the above Helmholtz equation with a delta-function 'source', namely, $(\nabla^2 + k_{eff}^2)G(x) = \delta(x)$. Then, we could express $\phi(x,\omega)$ as an integral: $\phi(x,\omega) = \int G(x - x_o)Q(x_o)dx_o$, where $Q(x)$ does represent the inhomogeneous term, $-\frac{S(x,\omega)}{D}$, in Eq. (19). Substituting this relation into Eq. (19) yields: $(\nabla^2 + k_{eff}^2)\phi(x,\omega) = \int [(\nabla^2 + k_{eff}^2)G(x - x_o)]Q(x_o)dx_o = \int \delta(x - x_o)Q(x_o)dx_o = Q(x)$. The $G(x - x_o)$ is thus called the Green function for the Helmholtz equation and represents the 'response' to a delta function source since $(\nabla^2 + k_{eff}^2)G(x) = \delta(x)$.

The derivation of the Green function for this standard Helmholtz equation [i.e., $(\nabla^2 + k_{eff}^2)\phi(x,\omega) = Q(x)$] requires very lengthy and complicated mathematical manipulations[34]. But the resulting Green function is rather simple[34,35]: $G(x) = -\frac{e^{ik_{eff} x}}{4\pi x}$. The response kernel $K(x)$ for the telegraph-type equation can be obtained by noting that the inhomogeneous term $Q(x)$ in the standard Helmholtz equation is replaced by the source term, $-\frac{S(x,\omega)}{D}$, in the Helmholtz equation relevant to the telegraph-type partial differential equation (PDE). Thus, the response kernel for the telegraph-type PDE is $K(x) = \left(-\frac{1}{D}\right)\left(-\frac{e^{ik_{eff}x}}{4\pi x}\right) = +\frac{e^{ik_{eff}x}}{4\pi Dx}$. It is damping Yukawa/ oscillatory-type kernel with the effective diffusion coefficient $D$ in the denominator. The corresponding Fourier-space Green function $K(k,\omega)$ can be obtained by carrying out the following four steps: (i) obtain the kernel $[\widetilde{K}(k,\omega) = (-Dk^2 + \tau\omega^2 - i\omega)^{-1}]$ by adopting $\phi(x,t) = \phi_o e^{-i(kx-\omega t)}$ in Eq. (8), (ii) take double Fourier transforms to obtain $\phi(k,\omega) = -\widetilde{K}(k,\omega)S(k,\omega)$, (iii) show that the source term $S(k,\omega)$ is directly proportional to $P(k,\omega)$ starting from $S(x,t) = \frac{1}{C_{eff}}[\nabla \cdot \boldsymbol{J}_{imp}(x,t)]$, and (iii) apply the linear convolution relation in $(k,\omega)$-space, $\phi(k,\omega) = K(k,\omega)P(k,\omega)$, to the resulting equation obtained in step (ii). Thus, the kernel in $(k,\omega)$-space is given by $K(k,\omega) = \frac{1}{+Dk^2-\tau\omega^2+i\omega}$.

**Acknowledgments**

H.M.J. acknowledges financial support from the National Research Foundation (NRF) grant funded by the Ministry of Science, ICT and Future Planning, Korea government (Grant No. NRF2020R1A2C1013915).

# Supplementary Information:

# Polarization-wave propagation as a biophysical mechanism of visual cognition

### Hyun Myung Jang,[1,2]* Youngwoo Jang,[3] and Hyeon Han[4]

[1]Research Institute of Advanced Materials, Seoul National University, Seoul 08826, Republic of Korea. [2](Permanent Address) Pohang University of Science and Technology (POSTECH), Pohang 37673, Republic of Korea. [3]Department of Cardiology, Gil Medical Center, Gachon University College of Medicine, Incheon 21565, Republic of Korea. [4]Department of Materials Science and Engineering, Pohang University of Science and Technology (POSTECH), Pohang 37673, Republic of Korea.

**\* Corresponds to   hmjang@postech.ac.kr**

## Supplementary Information includes:

**(i) Derivation of telegraph-type equation for cortical scalar potentials**

**(ii) Polarization wave packet and its dispersive spreading in time**

**(iii) Relation between $\nabla P(x,t)$ and $\nabla C(x,t)$ for a polarization wave packet**

# (i) Derivation of telegraph-type equation for cortical scalar potentials

In an isotropic, homogeneous cortical medium operating in the low-frequency regime, where the high-frequency component of the displacement current can be neglected, the divergence of the total current ($\nabla \cdot \boldsymbol{J}_{tot}$) vanishes, and the total current consists of four contributions: (i) conduction current, (ii) displacement current, (iii) polarization current, and (iv) impressed ionic current generated by neural activity. Accordingly, we write

$$\nabla \cdot \boldsymbol{J}_{tot} = 0, \quad \boldsymbol{J}_{tot} = \sigma \boldsymbol{E} + \varepsilon \frac{\partial \boldsymbol{E}}{\partial t} + \frac{\partial \boldsymbol{P}}{\partial t} + \boldsymbol{J}_{imp} \tag{S1}$$

where, $\sigma$ is the tissue conductivity, $\varepsilon$ the dielectric permittivity, $\boldsymbol{P}$ the polarization density, and $\boldsymbol{J}_{imp}$ the impressed ionic current. The electric field ($\boldsymbol{E}$) is given by

$$\boldsymbol{E} = -\nabla \phi \tag{S2}$$

Using Eq. (S2), the requirement of $\nabla \cdot \boldsymbol{J}_{tot} = 0$ in Eq. (S1) can be re-written as

$$-\sigma \nabla^2 \phi - \varepsilon \frac{\partial}{\partial t}(\nabla^2 \phi) + \frac{\partial}{\partial t}(\boldsymbol{\nabla} \cdot \boldsymbol{P}) + (\boldsymbol{\nabla} \cdot \boldsymbol{J}_{imp}) = 0 \tag{S3}$$

where $\boldsymbol{P}$ denotes the dielectric polarization which is defined as the dipole moment per unit volume.

Since the dipolar polarization in cortical tissue responds at low frequency, we adopt a Debye-type dielectric relaxation mechanism. Thus, we write

$$\tau_p \frac{\partial \boldsymbol{P}}{\partial t} + \boldsymbol{P} = \varepsilon_0 \chi \boldsymbol{E} = -\varepsilon_0 \chi \nabla \phi \tag{S4}$$

where $\chi$ is the effective dielectric susceptibility in the neuronal layer, $\varepsilon_0 = 8.854 \times 10^{-12} J^{-1}C^2m^{-1}$, the permittivity of free space, and $\tau_p$ the dipole-alignment relaxation time. Taking (i) time-derivative of the above equation and (ii) the divergence of the resulting equation, we obtain

$$\tau_p \frac{\partial^2}{\partial t^2}(\boldsymbol{\nabla} \cdot \boldsymbol{P}) + \frac{\partial}{\partial t}(\boldsymbol{\nabla} \cdot \boldsymbol{P}) = -\varepsilon_p \frac{\partial}{\partial t}(\nabla^2 \phi) \tag{S5}$$

where $\varepsilon_p$ denotes $\varepsilon_0 \chi$. Then, combining Eq. (S5) with Eq. (S3) to eliminate $\frac{\partial}{\partial t}(\boldsymbol{\nabla} \cdot \boldsymbol{P})$-term yields

$$-\sigma \nabla^2 \phi - \varepsilon \frac{\partial}{\partial t}(\nabla^2 \phi) - \tau_p \frac{\partial^2}{\partial t^2}(\boldsymbol{\nabla} \cdot \boldsymbol{P}) - \varepsilon_p \frac{\partial}{\partial t}(\nabla^2 \phi) + (\boldsymbol{\nabla} \cdot \boldsymbol{J}_{imp}) = 0 \tag{S6}$$

Under the single-mode *ansatz*, $P(x,t) = P_o e^{-i(kx-\omega t)} e^{-i\delta}$, where $\delta$ denotes the phase lag or retardation. Thus, $\tan \delta = \chi''(\omega)/\chi'(\omega)$, where $\chi''(\omega)$ denotes the imaginary part of the $\omega$-dependent dielectric susceptibility. Since the dipolar polarization in cortical tissue responds at low frequency, $e^{-i\delta} \approx e^0 = 1$. Therefore, we write the following expression for the time-dependent dielectric polarization:

$$P(x,t) = P_o e^{-i(kx-\omega t)} = P(t)e^{-ikx} \qquad (S7)$$

where $P(t) = P_o e^{+i\omega t}$. Similarly, we write

$$\phi(x,t) = \phi_o e^{-i(kx-\omega t)} = \phi(t)e^{-ikx} \qquad (S8)$$

where $\phi(t) = \phi_o e^{+i\omega t}$. In cortical medium, the dielectric permittivity $\varepsilon$ can be written as $\varepsilon_o \varepsilon_r$, where $\varepsilon_r$ denotes the relative dielectric permittivity (i.e., dielectric constant). Thus, we write $\varepsilon = \varepsilon_o \varepsilon_r = \varepsilon_o(1+\chi)$. If we assume that $\chi$ in cortical medium is substantially bigger than 1, we have the following approximation: $\varepsilon \approx \varepsilon_0 \chi$. Since $\varepsilon_p$ is defined as $\varepsilon_0 \chi$, we have the following approximation: $\varepsilon \approx \varepsilon_p = \varepsilon_0 \chi$. Substituting this expression of $\varepsilon_p$ into Eq. (S6), we obtain the following partial differential equation (PDE):

$$-\sigma \nabla^2 \phi - 2\varepsilon \frac{\partial}{\partial t}(\nabla^2 \phi) - \tau_p \frac{\partial^2}{\partial t^2}(\nabla \cdot \boldsymbol{P}) + (\nabla \cdot \boldsymbol{J}_{imp}) = 0 \qquad (S9)$$

where $\varepsilon \approx \varepsilon_p = \varepsilon_0 \chi$.

We now eliminate $\frac{\partial^2}{\partial t^2}(\nabla \cdot \boldsymbol{P})$ term in Eq. (S9) and express this in terms of $\frac{\partial^2}{\partial t^2}\phi(x,t)$. To do this, we first compute the divergence term, $\nabla \cdot \boldsymbol{P}$, using Eq. (S7): $\nabla \cdot \boldsymbol{P} = \frac{\partial}{\partial x}P(x,t) + \frac{\partial}{\partial y}P(x,t) + \frac{\partial}{\partial z}P(x,t)$. Since $\frac{\partial}{\partial y}P(x,t) = \frac{\partial}{\partial z}P(x,t) = 0$, $\nabla \cdot \boldsymbol{P} = \frac{\partial}{\partial x}P(x,t) = -kP(x,t)$. Using this expression, we calculate $\frac{\partial^2}{\partial t^2}(\nabla \cdot \boldsymbol{P})$ term to yield

$$\frac{\partial^2}{\partial t^2}(\nabla \cdot \boldsymbol{P}) = (-k)(+i\omega)^2 P_o e^{-i(kx-\omega t)} = +k\omega^2 P_o e^{-i(kx-\omega t)} \qquad (S10)$$

Similarly, we obtain

$$\frac{\partial^2}{\partial t^2}\phi(x,t) = -\omega^2 \phi_o e^{-i(kx-\omega t)} \qquad (S11)$$

Combining Eq. (S10) with Eq. (S11) immediately yields

$$\frac{\partial^2}{\partial t^2}(\nabla \cdot \boldsymbol{P}) = -k\left(\frac{P_o}{\phi_o}\right)\frac{\partial^2}{\partial t^2}\phi(x,t) \qquad (S12)$$

Similarly, $\frac{\partial}{\partial t}(\nabla^2 \phi)$ term in Eq. (S9) can be rewritten as

$$\frac{\partial}{\partial t}(\nabla^2 \phi) = -k^2 \frac{\partial \phi}{\partial t} \qquad (S13)$$

Substituting Eq. (S12) and Eq. (S13) into Eq. (S9) and rearranging yield

$$\tau_p k \left(\frac{P_o}{\phi_o}\right)\frac{\partial^2 \phi}{\partial t^2} + 2\varepsilon_0 \chi k^2 \frac{\partial \phi}{\partial t} = +\sigma \nabla^2 \phi - \nabla \cdot \boldsymbol{J}_{imp} \qquad (S14)$$

Dividing both sides of Eq. (S14) by $2\varepsilon_0\chi k^2$, we obtain

$$\tau \frac{\partial^2 \phi}{\partial t^2} + \frac{\partial \phi}{\partial t} = D\nabla^2 \phi + S(x,t) \tag{S15}$$

This is the telegraph-type PDE (Eq. (8) of the main manuscript) used in the simulation of the space-time map of the cortical scalar potential field $\phi(x,t)$. $\tau$ appeared in Eq. (S15) is called the effective relaxation time. $D$ denotes the effective diffusion coefficient in the cortical medium, and $S(x,t)$ is called the source term. They are defined as

$$\tau \equiv \frac{\tau_p\left(\frac{P_o}{\phi_o}\right)}{2\varepsilon_0 \chi k} = \frac{k\tau_p\left(\frac{P_o}{\phi_o}\right)}{C_{eff}} \tag{S16}$$

$$D = \frac{\sigma}{2\varepsilon_0 \chi k^2} = \frac{\sigma}{C_{eff}} \tag{S17}$$

And
$$S(x,t) = \frac{|\nabla \cdot J_{imp}|}{2\varepsilon_0 \chi k^2} = \frac{|\nabla \cdot J_{imp}|}{C_{eff}} \tag{S18}$$

where $C_{eff}$ is the effective capacitance factor and is defined as $2\varepsilon_0\chi k^2$.

### (ii) Polarization wave packet and its dispersive spreading in time

The propagating polarization field in the V1 region can be represented by a wave packet $P(x,t)$ composed of multiple $(k,\omega)$-modes distributed around the cognitively fundamental modulation parameters $(k_o,\omega_o)$. We thus write $P(x,t)$ by the following integral form using the $k$-dependent polarization $P(k)$:

$$P(x,t) = \int dk\, P(k)\, e^{-i\{kx-\omega(k)t\}} \tag{S19}$$

If we assume a Gaussian normal distribution, $P(k)$ can be written as

$$P(k) = P_o e^{-\alpha(k-k_o)^2} \tag{S20}$$

where $P_o \equiv P(k_o)$ and $\alpha \equiv 1/2\sigma^2$ with $\sigma$ defined as the standard deviation in the distribution of wave numbers ($ks$). In Eq. (S19), $\omega(k)$ denotes $k$-dependent angular frequency. It is described by Eq. (13) of the main manuscript. Substituting Eq. (S20) and Eq. (13) into Eq. (S19) yields

$$P(x,t) = P_o e^{-i\{k_o x - \omega_o t\}} \int_{-\infty}^{+\infty} dk'\, e^{-ik'(x-v_g t)} e^{-\left(\alpha - \frac{i\beta t}{2}\right)k'^2} \tag{S21}$$

where $k'$ is defined as $k' \equiv k - k_o$. On the other hand, we obtain the following expression for $P(x,0)$ from Eq. (S19):

$$P(x,0) \equiv P(x) = \int_{-\infty}^{+\infty} dk\, P(k) e^{-ikx} \tag{S22}$$

Substituting Eq. (S20) into Eq. (S22) yields

$$P(x) = P_o e^{-ik_o x} \int_{-\infty}^{+\infty} dk'\, exp\left[-\alpha\left\{k' + \left(\frac{ix}{2\alpha}\right)\right\}^2\right] exp\left(-\frac{x^2}{4\alpha}\right) \tag{S23}$$

It is known that $\int_{-\infty}^{+\infty} dk\, e^{-\alpha k^2} = \sqrt{\frac{\pi}{\alpha}}$. Using this equality, Eq. (S23) becomes

$$P(x) \equiv P(x,0) = \sqrt{\frac{\pi}{\alpha}} P_o e^{-ik_o x} e^{-x^2/4\alpha} \tag{S24}$$

Thus, we have

$$|P(x,0)| = \sqrt{\frac{\pi}{\alpha}} P_o e^{-x^2/4\alpha} \tag{S25}$$

Eq. (S24) can be rewritten as

$$P(x) = \sqrt{\frac{\pi}{\alpha}} P_o e^{-ik_o x} e^{-x^2/4\alpha} = P_o e^{-ik_o x} \int_{-\infty}^{+\infty} dk'\, e^{-ik'x} e^{-\alpha k'^2} \tag{S26}$$

By carefully comparing the second expression of Eq. (S26) with Eq. (S21) we deduce the following mathematical manipulations to obtain the final form of $P(x,t)$ from the first expression of Eq. (S26): (i) substitution of '$x - v_g t$' for '$x$' term appeared in $e^{-x^2/4\alpha}$ of Eq. (S26) and (ii) substitution of '$\alpha - \frac{i\beta t}{2}$' for '$\alpha$' term appeared in $(\pi/\alpha)^{1/2}$ and $e^{-x^2/4\alpha}$ of Eq. (S26). Thus, we obtain

$$P(x,t) = P_o e^{-i\{k_o x - \omega_o t\}} \left(\frac{\pi}{\alpha - \frac{i\beta t}{2}}\right)^{1/2} exp\left\{-\frac{(x-v_g t)^2}{4\left(\alpha - \frac{i\beta t}{2}\right)}\right\} \tag{S27}$$

This is the final mathematical expression of the polarization wave packet propagating with the group velocity of $v_g$ in the V1 region. To obtain the envelope $|P(x,t)|$ we should evaluate the magnitude of the following two terms that appeared in Eq. (S27):

$$\left|\left(\frac{\pi}{\alpha - \frac{i\beta t}{2}}\right)^{1/2}\right|^2 = \left(\frac{\pi}{\alpha - \frac{i\beta t}{2}}\right)^{1/2} \left\{\left(\frac{\pi}{\alpha - \frac{i\beta t}{2}}\right)^{1/2}\right\}^* = \pi^{1/2}\left\{\frac{\left(\alpha + \frac{i\beta t}{2}\right)}{\left(\alpha^2 + \frac{\beta^2 t^2}{4}\right)}\right\}^{1/2} \pi^{1/2}\left\{\frac{\left(\alpha - + \frac{i\beta t}{2}\right)}{\left(\alpha^2 + \frac{\beta^2 t^2}{4}\right)}\right\}^{1/2} \tag{S28}$$

$$\left|exp\left\{-\frac{(x-v_g t)^2}{4\left(\alpha - \frac{i\beta t}{2}\right)}\right\}\right|^2 = exp\left\{-\frac{\alpha(x-v_g t)^2}{2\left(\alpha^2 + \frac{\beta^2 t^2}{4}\right)}\right\} \tag{S29}$$

We obtain the following equation for the $x$-dependent envelope of the polarization wave packet by substituting these two relations into Eq. (S27):

$$|P(x,t)| = P_0 \left|e^{-i(k_o x - \omega_o t)}\right| \left(\frac{\pi^2}{\alpha^2 + \beta^2 t^2/4}\right)^{1/4} exp\left[-\frac{\alpha(x-v_g t)^2}{4(\alpha^2 + \beta^2 t^2/4)}\right] \tag{S30}$$

where $P_0 = P(k_o)$ and $|e^{-i(k_o x - \omega_o t)}| = 1$.

We now find the propagation velocity of $\phi(x,t)$ ridge. For this, we first consider the $k$-dependent scalar potential field under the condition of Gaussian normal distribution

$$\phi(k) = \phi_o e^{-\alpha(k-k_o)^2} \tag{S31}$$

Thus, $\phi(k,t)$ can be written as

$$\phi(k,t) = \phi(k)e^{+i\omega(k)t} = \phi_o e^{-\alpha(k-k_o)^2} e^{+i\omega(k)t} \tag{S32}$$

Similarly, $P(k,t)$ can be written using Eq. (S20) as

$$P(k,t) = P(k)e^{+i\omega(k)t} = P_o e^{-\alpha(k-k_o)^2} e^{+i\omega(k)t} \tag{S33}$$

Performing inverse Fourier transform, we get: $P(x,t) = \int dk\, P(k,t) e^{-ikx} = \int dk\, P(k) e^{-i\{kx-\omega(k)t\}}$. As expected, this is equivalent to Eq. (S19) presented previously. Under the single-mode ansatz, $P(x,t) = P_o e^{-i\{k_o x - \omega_o t\}}$. Then, $P(k,t)$ can be written using forward Fourier transform of $P(x,t)$, namely,

$$P(k,t) \equiv \int P(x,t) e^{+ikx} dx = P_o e^{+i\omega_o t} \int e^{-i(k_o-k)x} dx = 2\pi P_o e^{+i\omega_o t} \delta(k_o - k) \tag{S34}$$

The last expression of Eq. (S34) was obtained by applying the following equality: $2\pi \delta(k_o - k) = \int_{-\infty}^{+\infty} e^{-i(k_o-k)x} dx$. We have the linear convolution relation in $k$-space:

$$\phi(k,t) = K(k)P(k,t) \tag{S35}$$

This is Eq. (11) of the main manuscript. In the below, we will obtain this convolution relation starting from the convolution integral in $(x,t)$-space. Substituting Eq. (S34) into Eq. (S35) for $P(k,t)$ gives $\phi(k,t)$. Taking inverse Fourier transform of $\phi(k,t)$ then yields $\phi(x,t)$, namely,

$$\phi(x,t) = \int \phi(k,t) e^{-ikx} dk = 2\pi P_o e^{+i\omega_o t} \int K(k)\, \delta(k_o - k) e^{-ikx} dk = 2\pi P_o e^{+i\omega_o t} K(k_o) e^{-ik_o x} \equiv \phi_o e^{-i(k_o x - \omega_o t)} \tag{S36}$$

Thus, we deduce $K(k_o)$ from Eq. (S36)

$$K(k_o) = \frac{\phi_o}{(2\pi P_o)} \neq 0 \tag{S37}$$

Thus, $K(k)$ is a non-zero constant. On the other hand, $K(x)$ is given by $K(x) = \int K(k) e^{-ikx} dk = K(k)\, 2\pi\delta(x)$. The last expression was obtained by noting that $\int e^{-ikx} dk = 2\pi\delta(x)$. Thus, $K(x) = 2\pi K(k) = \frac{\phi_o}{P_o} \neq 0$ only when $x = 0$. Otherwise, $K(x) = 0$. According to the convolution relation, $\phi(x,t)$ can be written as

$$\phi(x,t) = \int K(x-x')P(x',t)\,dx' = \int K\big((x-vt)-x'\big)P(x',t)\,dx'$$

$$= (K * P)(x-vt) \tag{S38}$$

The second expression of the convolution integral is written for the travelling polarization wave with the velocity $v$. Accordingly, $P(x,t) = P(x-vt)$. Eq. (38) expresses the velocity preservation. Here, the kernel $K$ does not alter the propagation speed of the travelling polarization wave. Thus, $\phi(x,t)$ inherits the same propagation velocity as $P(x,t)$. The kernel $K(x-x')$ gives a non-zero contribution of $(\phi_o/P_o)$ only when $x' = x$. In other words, we establish

$$\phi(x,t) = K(0)P(x,t) = \left(\frac{\phi_o}{P_o}\right)P(x,t) \tag{S39}$$

Applying forward Fourier transform to Eq. (S39), we obtain the linear convolution relation in $k$-space, $\phi(k,t) = K(k)P(k,t)$ [Eq. (S35)]: $\phi(k,t) = \int \phi(x,t)\,e^{+ikx}dx = \int K(x-x')e^{+ik(x-x')}\,dx \int P(x',t)e^{+ikx'}\,dx' = K(k)P(k,t)$.

According to Eq. (S37), a non-zero value of $K(k)$ exists, which indicates that the propagation velocity of $\phi(x,t)$ ridge is equal to the group velocity of $P(x,t)$ as we deduce this equal speed from the velocity preservation in Eq. (S38). We now find the propagation velocity of $\phi(x,t)$ ridge to examine whether this prediction holds for $\phi(x,t)$ having a wave-packet form. To do this, we firstly compute $\phi(x,t)$ by applying inverse Fourier transform of $\phi(k,t)$ and Eq. (S32), namely,

$$\phi(x,t) = \int dk\,\phi(k,t)\,e^{-ikx} = \int dk\,\phi(k)\,e^{-i\{kx-\omega(k)t\}} \tag{S40}$$

We can obtain the following expression of $|\phi(x,t)|$ by repeatedly applying the procedures between Eq. (S19) and Eq. (S30) to the above expression of $\phi(x,t)$.

$$|\phi(x,t)| = \phi(k_o)\left(\frac{\pi^2}{\alpha^2+\beta^2t^2/4}\right)^{1/4}\exp\left[-\frac{\alpha(x-v_gt)^2}{4(\alpha^2+\beta^2t^2/4)}\right] \tag{S41}$$

According to Eq. (S41), the propagation velocity of wave-packet-like $\phi(x,t)$ ridge is equal to $v_g\,(=\frac{x}{t})$ because the exponential term in Eq. (S41) should be 1 at the $\phi(x,t)$ ridge. Thus, $(x-v_gt)^2$ appeared in the numerator of exponential term should be 0, showing that the propagation velocity of $\phi(x,t)$ ridge is equal to the group velocity of $P(x,t)$.

Because of dispersive spreading, the peak amplitude of the polarization wave packet decreases, and its width increases over time (Fig. 4B). To quantify this, let the peak intensity at time $t = 0$ and position $x = 0$ be denoted: $I_{0,0}$. Similarly, let the peak intensity at time $t = \Delta t$ and position $x = \Delta x$ be $I_{\Delta x, \Delta t}$. Using Eq. (S30), the ratio of these two intensities becomes:

$$\frac{I_{\Delta x,\Delta t}}{I_{0,0}} = \left(\frac{|P(\Delta x,\Delta t)|}{|P(0,0)|}\right)^2 = \left(\frac{\alpha^2}{\alpha^2+\beta^2(\Delta t)^2/4}\right)^{1/2} \tag{S42}$$

Since $\Delta x = v_g \Delta t$, the intensity decays entirely as a function of time. If we define the lifetime of the polarization wave packet as the time at which the peak intensity decays to one-tenth of its initial value, we denote this time as $t_{lf(p)}$. From Eq. (S42), we then obtain

$$t_{lf(p)} = \frac{(2\sqrt{99})\alpha}{\beta} \approx \frac{20\alpha}{\beta} \tag{S43}$$

Here, $\sigma = 1/\sqrt{2\alpha}$ is the standard deviation of the wave-number distribution; thus, larger $\sigma$ (broader $k$-distribution) leads to a shorter lifetime $t_{lf(p)}$. The parameter $\beta = (d^2\omega/dk^2)_{k_0}$ is the principal factor controlling the spreading of the wave packet; smaller $\beta$ yields a longer lifetime.

In addition to amplitude decay, the lifetime can also be defined through the spreading of the half-width at half-maximum (HWHM) of the Gaussian envelope. Let the HWHM at $t = 0$ be $W_d(0)$. Similarly, let the HWHM at $t = \Delta t$ be $W_d(\Delta t)$. From Eq. (S30), we obtain the following expression for $|P(x,t)|$ at $t = 0$:

$$|P(x,0)| = P_o \left(\frac{\pi}{\alpha}\right)^{1/2} e^{-x^2/4\alpha} \tag{S44}$$

We then consider the time-dependent HWHM to show the degree of spreading in time. According to the definition of HWHM, we have the following requirement at $t = 0$:

$$|P(x_{HM},0)| = \frac{1}{2}|P(0,0)| \tag{S45}$$

where $x_{HM}$ denotes the position of HWHM at $t = 0$. Thus, $x_{HM} = W_d(0)$. Combining Eq. (S45) with Eq. (S44) yields

$$W_d(t=0) = W_d(0) = x_{HM} = \sqrt{4\alpha \ln 2} \tag{S46}$$

The peak position of $P(x,t)$ at time $t = \Delta t$ is $\Delta x = v_g \Delta t$. Thus, we have the following expression for $|P(\Delta x, \Delta t)|$ from Eq. (S30):

$$|P(\Delta x, \Delta t)| = P_0 \left(\frac{\pi^2}{\alpha^2+\beta^2(\Delta t)^2/4}\right)^{1/4} \tag{S47}$$

On the contrary, we have the following expression of $|P(x_{HM}, \Delta t)|$ at $x = x_{HM}$ from Eq. (S30):

$$|P(x_{HM}, \Delta t)| = P_0 \left(\frac{\pi^2}{\alpha^2 + \beta^2(\Delta t)^2/4}\right)^{1/4} \exp\left[-\frac{\alpha(x_{HM} - v_g \Delta t)^2}{4\{\alpha^2 + \frac{\beta^2(\Delta t)^2}{4}\}}\right] \tag{S48}$$

According to the definition of HWHM, we establish the following equality:

$$\frac{1}{2}|P(\Delta x, \Delta t)| = |P(x_{HM}, \Delta t)| \tag{S49}$$

Substituting Eq. (S47) and Eq. (S48) into the equality requirement given in Eq. (S49) yields the following relation at $x = x_{HM}$:

$$4\left(\alpha^2 + \frac{\beta^2(\Delta t)^2}{4}\right) \ln 2 = \alpha(x_{HM} - v_g \Delta t)^2 \tag{S50}$$

On the other hand, as shown in the $|P(x,t)|$ profile presented in the Figure below, we have the following expression for $x_{HM}$ at $t = \Delta t$:

$$x_{HM} = HWHM + \Delta x = W_d(\Delta t) + v_g \Delta t \tag{S51}$$

Substituting Eq. (S51) into Eq. (S50) and comparing this with Eq. (S46) yields

$$\frac{W_d(\Delta t)}{W_d(0)} = \left(1 + \frac{\beta^2(\Delta t)^2}{4\alpha^2}\right)^{1/2} \tag{S52}$$

This is Eq. (18) of the main manuscript. If the lifetime is defined as the time when the HWHM (i.e., $W_d$) becomes ten times broader than its initial value, one obtains: $t_{lf(w)} = \frac{(2\sqrt{99})\alpha}{\beta} \approx \frac{20\alpha}{\beta}$. The remarkable fact that two distinct definitions of lifetime yield exactly the same functional form $t_{lf} = (2\sqrt{99})\alpha/\beta$ highlights the internal mathematical consistency of the spreading behavior of $|P(x,t)|$ in the V1 region.

We can also deduce the uncertainty relation by examining two profiles, namely, (i) $k$-dependent polarization profile $P(k)$ and (ii) the polarization wave packet $P(x,t)$ before undergoing the dispersive spreading due to non-zero $\beta$. If we take the uncertainty in the wave number $\Delta k$ as the HWHM of the $P(k)$ profile, we then establish the following equality from Eq. (S20): $P_o e^{-\alpha(k_o + \Delta k - k_o)^2} = \frac{1}{2} P_o e^{-0}$. This immediately yields the following expression for the uncertainty in the wave number $\Delta k$:

$$\Delta k = \sqrt{\frac{\ln 2}{\alpha}} \tag{S53}$$

As shown in Eq. (46), the uncertainty ($\Delta x$) in the position of the polarization wave packet $P(x,t)$ at $t = 0$ is $\Delta x\,(= x_{HM}) = \sqrt{4\alpha\,ln2}$. Thus, we obtain the following ($\Delta k - \Delta x$) uncertainty product:

$$\hbar\,\Delta k \Delta x = 2ln2\,\hbar = 1.38\,\hbar \geq \frac{1}{2}\hbar \tag{S54}$$

Therefore, Heisenberg's uncertainty relation also holds for the polarization wave packet in the V1 region.

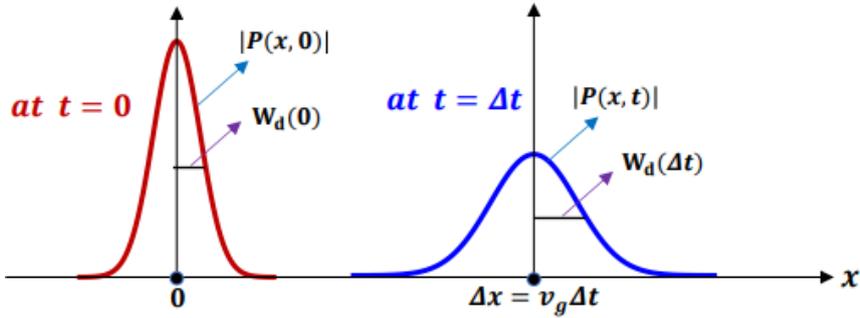

**(Figure Caption)** Propagation of the polarization wave packet $|P(x,t)|$ with the group velocity $v_g$, showing spreading behavior owing to non-zero $\beta$. Initially, the peak value of $|P(x,0)|$ at $x = 0$ is given by $|P(0,0)| = \left(\frac{\pi}{\alpha}\right)^{\frac{1}{2}} P_o$ with its HWHM denoted by $W_d(0) = \sqrt{4\alpha \cdot ln2}$. On the contrary, the peak value of $|P(x,t)|$ at $t = \Delta t$ is given by $|P(\Delta x, \Delta t)| = P_o \left(\frac{\pi^2}{\alpha^2 + \frac{\beta^2 (\Delta t)^2}{4}}\right)^{\frac{1}{4}}$, where $\Delta x = v_g \Delta t$. As shown in Eq. (S52), $W_d(\Delta t)$ is given by the following expression of the dispersive spreading: $W_d(\Delta t) = W_d(0)\left(1 + \frac{\beta^2 (\Delta t)^2}{4\alpha^2}\right)^{\frac{1}{2}}$.

### (iii) Relation between $\nabla P(x,t)$ and $\nabla C(x,t)$ for a polarization wave packet

In Eq. (15) of the main manuscript, we showed the following expression for a polarization wave packet:

$$P(x,t) = P(k_0)\,e^{-i(k_0 x - \omega_0 t)} A(t) \exp\left[-\frac{(x - v_g t)^2}{4\left(\alpha - \frac{i\beta t}{2}\right)}\right] \tag{15}$$

The corresponding scalar potential field can be expressed as

$$\phi(x,t) = \phi(k_0)e^{-i(k_0 x - \omega_0 t)}A(t)\exp\left[-\frac{(x-v_g t)^2}{4\left(\alpha-\frac{i\beta t}{2}\right)}\right] \tag{S55}$$

We can readily find the following expression of $\nabla P(x,t)$ from Eq. (15):

$$\nabla P(x,t) = P(k_0)e^{-i(k_0 x - \omega_0 t)}A(t)\exp\left[-\frac{(x-v_g t)^2}{4\left(\alpha-\frac{i\beta t}{2}\right)}\right]\left\{(-ik_o) - \frac{(x-v_g t)}{2\left(\alpha-\frac{i\beta t}{2}\right)}\right\} \tag{S56}$$

Similarly, we have the following expression of $\nabla\phi(x,t)$ from Eq. (S55):

$$\nabla\phi(x,t) = \phi(k_0)e^{-i(k_0 x - \omega_0 t)}A(t)\exp\left[-\frac{(x-v_g t)^2}{4\left(\alpha-\frac{i\beta t}{2}\right)}\right]\left\{(-ik_o) - \frac{(x-v_g t)}{2\left(\alpha-\frac{i\beta t}{2}\right)}\right\} \tag{S57}$$

We thus have the following ratio from Eqs. (S56) and (S57):

$$\frac{\nabla P(x,t)}{\nabla\phi(x,t)} = \frac{P(k_o)}{\phi(k_o)} \equiv \frac{P_o}{\phi_o} = \frac{1}{2\pi K(k_o)} \tag{S58}$$

The last expression of Eq. (S58) is obtained from Eq. (S37).

We have the following equality from Fick's law and Ohm's law:

$$\boldsymbol{J}_{con} = -D\nabla C(x,t) = \sigma\boldsymbol{E} = -\sigma\nabla\phi(x,t) \tag{S59}$$

This macroscopic phenomenological relation is valid, regardless of the number of $(k,\omega)$ components in a system. Combining Eq. (S58) with Eq. (S59) to eliminate $\nabla\phi(x,t)$ term yields

$$\nabla P(x,t) = \frac{1}{2\pi K(k_0)}\left(\frac{D}{\sigma}\right)\nabla C(x,t) \tag{S60}$$

Eq. (S60) is equal to the first expression of Eq. (14) of the main manuscript. As noted in the main manuscript, Fick's first law in Eq. (S59) can be rewritten in terms of the chemical potential gradient as

$$D\nabla C(x,t) = -\boldsymbol{J}_{con} = +bC(x,t)\nabla\mu(x,t) \tag{S61}$$

This macroscopically valid relation also holds for a multi-$(k,\omega)$ components system. Substituting Eq. (S61) into Eq. (S60) yields

$$\nabla P(x,t) = \frac{1}{2\pi K(k_0)}\left(\frac{b}{\sigma}\right)C(x,t)\,\nabla\mu(x,t) \tag{S62}$$

Eq. (S62) is equal to the second expression of Eq. (14) for $\nabla P(x,t)$.